\RequirePackage{fix-cm}
\documentclass{article}
\usepackage{spconf,amsmath,amssymb,graphicx,hyperref,booktabs,tabularx,xcolor}
\hypersetup{hidelinks}
\makeatletter
\renewcommand\section{\@startsection{section}{1}{\z@}%
  {-2.5ex plus -.4ex minus -.2ex}{1.2ex plus .2ex}{\normalfont\bfseries}}
\renewcommand\subsection{\@startsection{subsection}{2}{\z@}%
  {-2.0ex plus -.3ex minus -.2ex}{.9ex plus .2ex}{\normalfont\bfseries}}
\newenvironment{fixedfigure}{\par\noindent\begin{minipage}{\columnwidth}%
  \small\def\@captype{figure}}{\end{minipage}\par\vspace{6pt}}
\newenvironment{fixedtable}{\par\noindent\begin{minipage}{\columnwidth}%
  \small\def\@captype{table}}{\end{minipage}\par\vspace{6pt}}
\newcommand{\widecaption}{\def\@captype{table}\caption}
\long\def\@makecaption#1#2{\vskip4pt
  \setbox\@tempboxa\hbox{#1. #2}%
  \ifdim\wd\@tempboxa>\hsize #1. #2\par
  \else\hbox to\hsize{\hfil\box\@tempboxa\hfil}\fi}
\makeatother
\newcommand{\heldout}{\textit{held-out subset}}

\title{WHAT MATTERS IN PREFERENCE POST-TRAINING OF CONTINUOUS AUTOREGRESSIVE SPEECH LANGUAGE MODELS}

\name{\begin{tabular}[t]{@{}c@{}}
    Yingda Shen$^{\dagger\star}$ \qquad Yao Qian$^{\star}$ \qquad Dekun Chen$^{\dagger}$ \qquad Yufei Liu$^{\star}$\\
    Junan Zhang$^{\dagger}$ \qquad Chong Jing$^{\dagger}$ \qquad Han Wang$^{\dagger}$ \qquad Yuxiang Wang$^{\dagger}$ \qquad Zhizheng Wu$^{\dagger}$
    \end{tabular}}
    
    \address{$^{\dagger}$ The Chinese University of Hong Kong, Shenzhen \\
    $^{\star}$ Microsoft Research}

\begin{document}
\maketitle
\fontsize{9.5}{11.2}\selectfont
\setlength{\abovedisplayskip}{4pt}
\setlength{\belowdisplayskip}{4pt}
\setlength{\abovedisplayshortskip}{2pt}
\setlength{\belowdisplayshortskip}{3pt}
\raggedbottom

% PAGE 1 -- LEFT COLUMN: abstract and introduction.
\begin{abstract}
Continuous autoregressive speech language models combine an autoregressive (AR) backbone with a diffusion or flow-matching head. Preference post-training requires choosing how to construct pairs and which modules to update. Controlled interventions show that a perturbed acoustic patch changes subsequent backbone states, yet larger representation changes do not necessarily reduce speech quality. We construct preference pairs through noise or inference-configuration perturbations and formulate a preference update based on reference-relative flow-matching errors. Updating the backbone with a fixed head yields stronger reward improvements than head-only adaptation in our evaluated setting. Across flow-time and module ablations, the lowest training loss does not consistently produce the best speech. These findings support preference post-training that considers both acoustic sampling and autoregressive conditioning, with training choices validated on generated utterances.
\end{abstract}
\begin{keywords}
Continuous speech language models, autoregressive generation, flow matching, preference optimization
\end{keywords}

\section{Introduction}
\label{sec:intro}

Continuous autoregressive speech language models (SLMs) represent speech as continuous acoustic patches and generate each patch with a diffusion or flow-matching head conditioned on an AR backbone~\cite{jia2025ditar,peng2025vibevoice,he2025continuoustoken}. Recent systems extend this approach to expressive synthesis, long-form speech, and unified generation and editing~\cite{inclusion2025ming,li2025dotstts,shen2026fireredtts3}. Post-training aims to make these capabilities reliable across inputs.

For a given input, generated utterances vary in intelligibility, speaker similarity, and perceptual quality. Preference alignment has improved speech synthesis and restoration by learning from such differences~\cite{zhang2024speechalign,tian2024preference,zhang2025intelligibility,zhang2025multimetric}. Direct Preference Optimization (DPO)~\cite{rafailov2023dpo} learns from paired responses, and prediction-error objectives extend this principle to diffusion-based generation~\cite{wallace2024diffusiondpo,liu2025ardmdpo}. The objective alone determines neither useful candidate comparisons nor which modules to update. We study candidate perturbation, flow-time sampling, and trainable modules as connected choices in continuous AR post-training.

These choices interact through autoregressive feedback. Useful preference supervision must account for both local acoustic sampling and its effects on subsequent generation.

\newpage
% PAGE 1 -- RIGHT COLUMN: architecture, contributions, generation model.
\begin{fixedfigure}
\centering
\includegraphics[width=.98\linewidth]{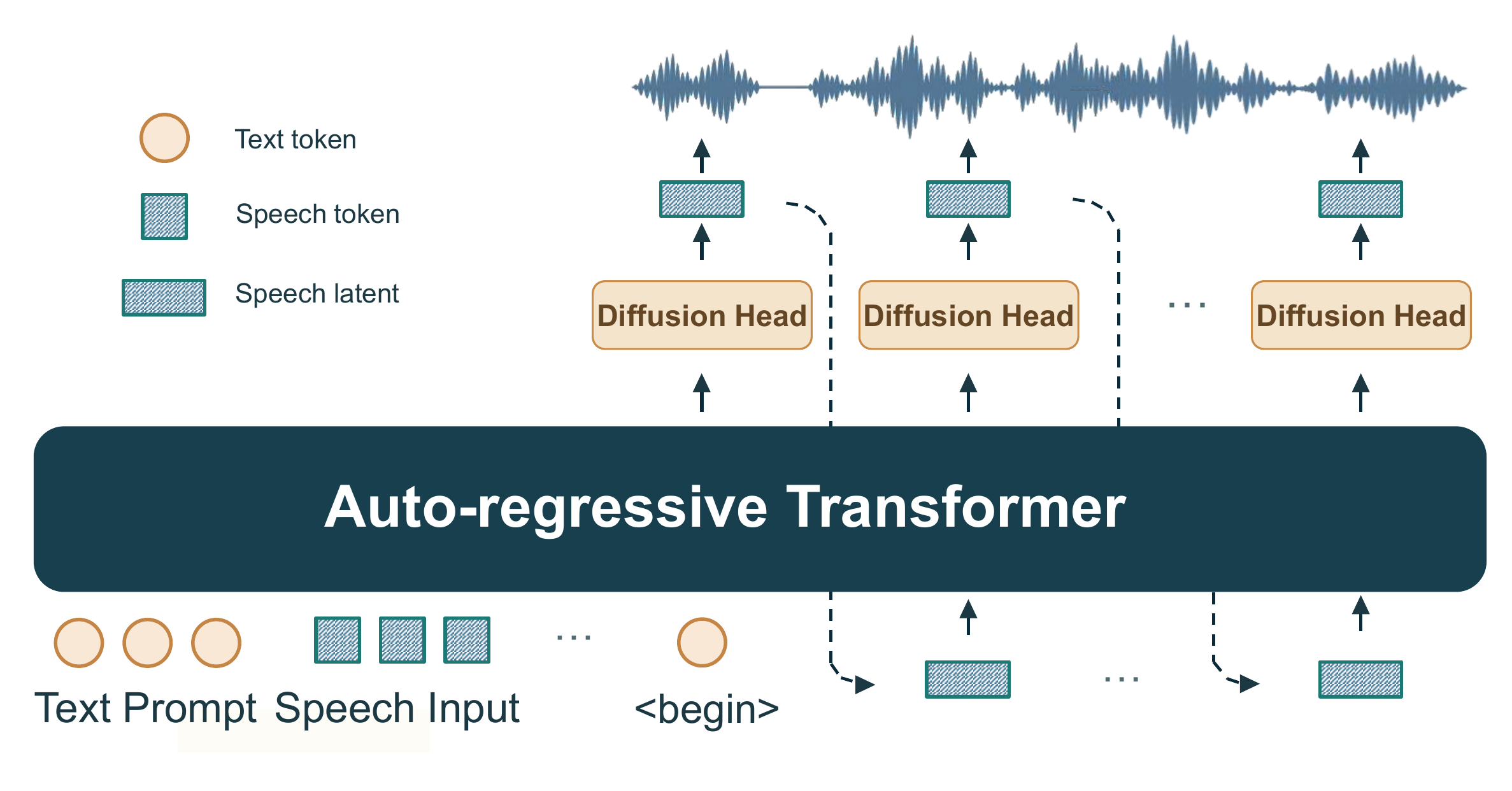}
\caption{Continuous autoregressive speech language model. The diffusion head generates acoustic patches conditioned on the AR backbone; generated patches become subsequent speech history.}
\label{fig:arch}
\end{fixedfigure}

Each sampled patch becomes input to the AR backbone, so a local sampling change can alter subsequent generation. Improving speech may therefore require adapting backbone conditions as well as acoustic sampling. We study both modules under the same preference objective.

We connect generation dynamics, candidate construction, and preference training through three contributions:
\begin{enumerate}
\setlength{\itemsep}{1pt}\setlength{\parsep}{0pt}\setlength{\topsep}{2pt}
\item Controlled replay and single-patch interventions separate local sampling effects from subsequent generation changes and relate them to speech quality.
\item Candidate rankings and reward margins motivate preference pairs from noise or inference-configuration perturbations.
\item Flow-time and module ablations show that backbone updates can improve speech with a fixed head, while training loss alone does not identify the best outputs.
\end{enumerate}

\section{Sampling and autoregressive feedback}
\label{sec:decouple}

Given input \(c\) and previous acoustic patches \(z_{<k}\), the backbone \(F_\phi\) and head \(G_\theta\) generate
\begin{equation}
\label{eq:loop}
h_k=F_\phi(c,z_{<k}),\qquad
z_k=G_\theta(\epsilon_k;h_k,z_{<k},a),
\end{equation}
where \(a\) specifies inference settings and \(\epsilon_k\) denotes sampling noise. Each new patch enters the history and changes subsequent conditions (Fig.~\ref{fig:arch}). We examine this feedback with three interventions.

\newpage
% PAGE 2 -- LEFT COLUMN: mechanism figure and Table 1.
\begin{fixedfigure}
\centering
\includegraphics[width=\linewidth]{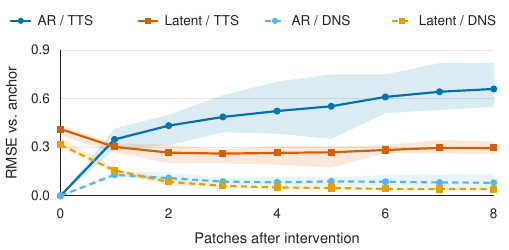}
\caption{Single-patch intervention at lag 0. RMSE is relative to the anchor; lines average over intervention positions and bands show their range.}
\label{fig:mech}
\end{fixedfigure}

\begin{fixedtable}
\centering
\caption{Noise perturbation with fixed or evolving conditions. Relative \(L_2\) is measured against the anchor; \(\Delta\)OVRL is the change in DNSMOS overall quality.}
\label{tab:internal}
\setlength{\tabcolsep}{3pt}
\begin{tabular}{@{}llccc@{}}
\toprule
Task & Setting & AR \(L_2\) & Latent \(L_2\) & \(\Delta\)OVRL \\
\midrule
TTS & frozen AR & \(0\) & \(0.95\) & \(-0.282\) \\
TTS & closed loop & \(0.65\) & \(1.22\) & \(+0.013\) \\
DNS & frozen AR & \(0\) & \(0.70\) & \(-0.050\) \\
DNS & closed loop & \(0.29\) & \(0.86\) & \(+0.022\) \\
\bottomrule
\end{tabular}
\end{fixedtable}

\textbf{Closed-loop rollout.} We change the sampling noise and let the model continue normally. Each new patch can affect subsequent backbone states, acoustic contexts, and stopping decisions.

\textbf{Frozen-trajectory replay.} We first record an anchor utterance, then resample its patches while keeping the anchor's backbone states, latent contexts, and stopping decisions fixed. Reusing the original noise reproduces the anchor. Unlike freezing backbone weights, this intervention prevents changed patches from influencing future conditions.

\textbf{Single-patch intervention.} We resample one patch and then resume normal autoregressive generation. We measure distances from the anchor starting at this intervention, excluding the unchanged prefix.

Table~\ref{tab:internal} reports results for text-to-speech (TTS) and speech denoising (DNS). Replay changes the acoustic latents while leaving backbone states unchanged; normal rollout changes both. The single-patch experiment shows how this happens over time (Fig.~\ref{fig:mech}): a latent change at the intervention is followed by changes in backbone states. Backbone divergence grows in TTS, whereas it decreases after an initial rise in DNS. The effect of a local perturbation therefore depends on the task.

Representation changes alone do not explain speech quality. Frozen replay reduces DNSMOS overall quality (OVRL), while normal rollout produces small positive changes despite larger latent distances. Allowing future conditions to change can therefore give a different outcome from resampling patches under fixed conditions. This does not establish an explicit correction mechanism, since replay may create inconsistent patch--context combinations. It does show why preferences should be evaluated on complete utterances and why both the backbone and the head are relevant to post-training.

\newpage
% PAGE 2 -- RIGHT COLUMN: preference figure and pair construction.
\begin{fixedfigure}
\centering
\includegraphics[width=\linewidth]{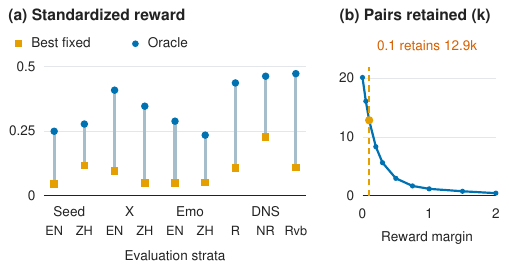}
\caption{Preference-pool analysis: (a) best fixed configuration versus per-prompt oracle reward; (b) comparisons retained by reward-margin filtering.}
\label{fig:pref}
\end{fixedfigure}

\section{Constructing preferences from generation variability}
\label{sec:supply}

We vary random inputs or inference settings and score each candidate \(y\) for input \(c\) with a fixed reward \(R(c,y)\). In a diagnostic pool of eight configurations, the most frequent winner is best on only \(19.4\%\) of prompts; normalized winner entropy ranges from \(0.830\) to \(0.977\). A per-prompt oracle gains \(0.16\)--\(0.36\) standardized reward units over the best fixed configuration (Fig.~\ref{fig:pref}a), showing that the preferred configuration varies across inputs.

Of \(20{,}160\) within-prompt comparisons, \(12{,}867\), \(8{,}366\), and \(2{,}978\) remain at reward margins of \(0.1\), \(0.2\), and \(0.5\), respectively (Fig.~\ref{fig:pref}b). These pool comparisons are not independent training pairs. Larger margins retain fewer, more distinct pairs.

\textbf{Noise perturbation (NP)} varies noise injected during stochastic differential equation (SDE) sampling at fixed inference settings. \textbf{Configuration perturbation (CP)} varies sampler settings, including ordinary differential equation (ODE) sampling, classifier-free guidance (CFG) strength, and the number of function evaluations (NFE), with shared initial noise. Pair selection and preference training are identical for NP and CP. Ming supports both; the ODE-only dots.tts head uses CP.

For each prompt, we pair the highest- and lowest-reward valid candidates when \(R(c,y^w)-R(c,y^l)\geq\delta\), using automatic reward labels.

\section{Preference optimization with flow matching}
\label{sec:probe}

\subsection{Comparing preferred and rejected speech}

DPO~\cite{rafailov2023dpo} favors the preferred response's log-density ratio to a reference over that of the rejected response. Since sequence likelihoods are costly, we derive a flow-matching surrogate from prediction errors~\cite{wallace2024diffusiondpo,liu2025ardmdpo}.

A variational autoencoder (VAE) encodes each waveform as \(z=(z_1,\ldots,z_K)\). At flow time \(t\), we interpolate between Gaussian noise and each clean patch~\cite{lipman2023flow}:
\begin{equation}
\label{eq:path}
x_{t,k}=(1-t)\epsilon_k+t z_k,\qquad
u_k=z_k-\epsilon_k,
\end{equation}
where \(\epsilon_k\sim\mathcal{N}(0,I)\) and \(u_k\) is the target velocity. Here, \(t=0\) denotes noise and \(t=1\) denotes clean data.

\newpage
% PAGE 3 -- LEFT COLUMN: time/scope training curves and preference loss.
\begin{fixedfigure}
\centering
\includegraphics[width=.94\linewidth]{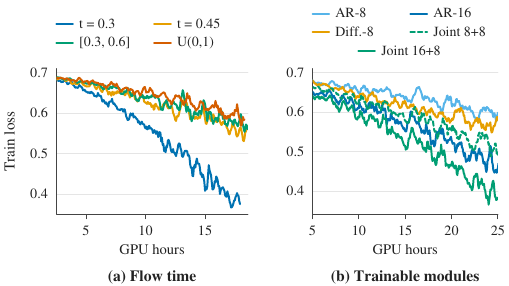}
\caption{Training curves for Table~\ref{tab:train}: (a) flow-time sampling (AR-8); (b) trainable scope.}
\label{fig:train}
\end{fixedfigure}

Let \(\psi=(\phi,\theta)\) and \(s_\psi(z)=\log[p_\psi(z\mid c)/p_0(z\mid c)]\), where \(p_0\) is the reference distribution. DPO uses the pair margin \(s_\psi(z^w)-s_\psi(z^l)\). To approximate these scores, we define the flow-matching energy
\begin{equation}
\label{eq:energy}
\begin{aligned}
E_{\phi,\theta}(z;c,t,\epsilon)
&=\frac{1}{Kd}\sum_{k=1}^{K}
\left\|v_\theta(x_{t,k},t;h_k)-u_k\right\|_2^2,\\
h_k&=F_\phi(c,z_{<k}).
\end{aligned}
\end{equation}
Here \(d\) is the patch dimensionality. The energy averages errors over valid patches under teacher-forced backbone conditions; additional head context is omitted from the notation.

The reference uses the same model with LoRA disabled and recomputes \(h_k^{\mathrm{ref}}=F_{\phi_0}(c,z_{<k})\) from the same clean latent history. Substituting negative energy for log density gives a reference-relative score \(\Delta(z)\) and pair margin
\begin{equation}
\label{eq:margin}
\begin{aligned}
\Delta(z)&=E_{\mathrm{ref}}(z)-E_{\phi,\theta}(z),\\
M_{\phi,\theta}&=\Delta(z^w)-\Delta(z^l).
\end{aligned}
\end{equation}
The training objective is
\begin{equation}
\label{eq:dpo}
\mathcal{L}_{\mathrm{pref}}
=-\mathbb{E}_{(c,z^w,z^l),\,t\sim p_t,\,\Xi}
\log\sigma\!\left(\beta M_{\phi,\theta}\right),
\end{equation}
where \(\sigma\) is the sigmoid function, \(\Xi=(\epsilon^w,\epsilon^l)\) denotes training noise, and \(\beta\) scales the margin. This unweighted flow surrogate approximates the likelihood-based objective; \(p_t\) is an explicit training choice. For a sampled pair, let \(\ell=-\log\sigma(\beta M)\) and \(E_\psi^{w/l}=E_{\phi,\theta}(z^{w/l})\). With the reference fixed,
\begin{equation}
\label{eq:update}
\nabla_\psi\ell=\beta\sigma(-\beta M)
\left(\nabla_\psi E_\psi^w-\nabla_\psi E_\psi^l\right).
\end{equation}
The update compares winner and loser gradients, weighting pairs with lower reference-relative margins more strongly.

\subsection{Time sampling and trainable modules}

Each pair shares one scalar \(t\) across both candidates and all patches. A fresh shared seed generates noise patch by patch: the candidates receive identical noise over their common patch range, while the longer candidate receives additional draws. For each candidate, the model and reference reuse the same noise and time.

\newpage
% PAGE 3 -- RIGHT COLUMN: Table 3, update choices, experimental setting.
\setcounter{table}{2}
\begin{fixedtable}
\centering
\caption{Held-out time and module ablations (95\% CI); WER is a fraction. Time: step 300; \([0.3,0.6]\) denotes continuous uniform sampling. Scope: best evaluated checkpoint.}
\label{tab:train}
\setlength{\tabcolsep}{1.5pt}
\small
\begin{tabular}{@{}lccccc@{}}
\toprule
\(t\) & \(\Delta\)rew. & 95\% CI & WER & SIM & DNSMOS \\
\midrule
Fixed \(0.3\) & \(+0.131\) & \([-0.072, 0.354]\) & \(0.124\) & \(0.454\) & \(2.85\) \\
Fixed \(0.45\) & \(+0.137\) & \([0.025, 0.280]\) & \(0.131\) & \(0.468\) & \(2.82\) \\
\([0.3,0.6]\) & \(+0.069\) & \([-0.089, 0.253]\) & \(0.137\) & \(0.473\) & \(2.84\) \\
\(U(0,1)\) & \(-0.069\) & \([-0.224, 0.079]\) & \(0.145\) & \(0.468\) & \(2.80\) \\
\midrule
Scope & \(\Delta\)rew. & 95\% CI & WER & SIM & DNSMOS \\
\midrule
AR last-8 & \(+0.470\) & \([0.181, 0.807]\) & \(0.089\) & \(0.488\) & \(2.87\) \\
AR last-16 & \(+0.737\) & \([0.361, 1.169]\) & \(0.060\) & \(0.508\) & \(2.88\) \\
Diff. last-8 & \(+0.076\) & \([-0.015, 0.187]\) & \(0.140\) & \(0.464\) & \(2.84\) \\
Joint 8{+}8 & \(+0.146\) & \([-0.000, 0.328]\) & \(0.128\) & \(0.460\) & \(2.88\) \\
Joint 16{+}8 & \(+0.701\) & \([0.334, 1.122]\) & \(0.063\) & \(0.509\) & \(2.91\) \\
\bottomrule
\end{tabular}
\end{fixedtable}
Our main experiments sample \(t\) uniformly from the discrete generation-schedule times in \([0.3,0.6]\). The noised inputs are newly constructed using~\eqref{eq:path}; neither the noise used to generate the preference pairs nor intermediate inference states are replayed.

We compare three update choices. \textbf{Head-only} training updates \(v_\theta\), with identical backbone conditions for the model and reference on a given candidate. \textbf{Backbone-only} training keeps the head weights fixed, but backpropagates through the head into \(h_k\) and the AR backbone. The model and reference then compute different conditions for the same candidate. \textbf{Joint} training updates both modules. In all cases, winner and loser conditions are computed from their respective histories. Training uses teacher forcing; evaluation generates complete utterances autoregressively.

\section{Experiments}
\label{sec:expts}

\subsection{Setting}
\label{ssec:setup}

Training prompts mix LibriTTS-R read speech and EARS emotional speech in a 1:1 ratio. Training pairs select the best and worst of six valid candidates, with a reward margin of at least \(0.1\). The reward combines Whisper-large-v3 word error rate (WER), WavLM speaker similarity, emotion2vec similarity, DNSMOS~\cite{radford2023whisper,chen2022wavlm,ma2025emotion2vec,reddy2021dnsmos}, duration, and validity. Our training uses \(\beta=200\), LoRA rank 8, constant learning rate \(5\times10^{-5}\), and an effective batch size of 128. Chosen-only supervised fine-tuning (SFT) trains on the winners. Both ARDM-DPO baselines update the last 16 AR layers (AR-16) with the published loss and training configuration~\cite{liu2025ardmdpo}, including \(t\sim U(0,1)\). For candidate generation, our baseline implementations vary only the initial Gaussian noise, using each model's native sampler: SDE for Ming and ODE for dots.tts.

The exploratory \heldout{} contains 100 English prompts; Ming-base scores reward \(7.44\), WER \(0.143\), SIM \(0.466\), and DNSMOS \(2.82\). The official CV3-Eval comparison (Table~\ref{tab:cv3}) uses 500 English zero-shot and 150 English emotion prompts. Some Ming-base generations collapse, producing WER above \(100\%\) and inflating the aggregate error rate. To assess cross-model transferability, we also evaluate our method on dots.tts, which uses a continuous AR architecture.

\clearpage
% PAGE 4 -- FULL-WIDTH TABLE: CosyVoice3, dots.tts, then Ming.
\twocolumn[{
\setcounter{table}{1}
\begin{minipage}{\textwidth}
\small\centering
\widecaption{Official CV3-Eval English results (WER in percent). \(\mathrm{SIM_E}\): ERes2Net (percent); \(\mathrm{SIM_W}\): WavLM cosine; \(\mathrm{SIM_{emo}}\): emotion2vec score-distribution cosine. NP/CP: noise/configuration perturbation.}
\label{tab:cv3}
\setlength{\tabcolsep}{3pt}
\begin{tabular*}{\textwidth}{@{\extracolsep{\fill}}lccccccccc@{}}
\toprule
& \multicolumn{4}{c}{Zero-shot / en} & \multicolumn{5}{c}{Emotion / en} \\
\cmidrule(lr){2-5}\cmidrule(lr){6-10}
System & WER\(\downarrow\) & \(\mathrm{SIM_E}\uparrow\) & \(\mathrm{SIM_W}\uparrow\) & OVRL\(\uparrow\) & WER\(\downarrow\) & \(\mathrm{SIM_E}\uparrow\) & \(\mathrm{SIM_W}\uparrow\) & OVRL\(\uparrow\) & \(\mathrm{SIM_{emo}}\uparrow\) \\
\midrule
CosyVoice3-0.5B & 5.44 & 74.0 & 0.675 & 3.12 & 5.04 & 66.8 & 0.576 & 3.15 & 0.631 \\
CosyVoice3-0.5B-RL & 3.74 & 73.7 & 0.657 & 3.15 & 5.74 & 65.7 & 0.569 & 3.14 & 0.591 \\
\midrule
dots.tts base & 5.22 & 76.1 & 0.732 & 3.06 & 7.17 & 69.3 & 0.666 & 2.90 & 0.641 \\
dots.tts ARDM-DPO (AR-16) & 5.33 & 75.9 & 0.732 & 3.06 & 7.01 & 69.6 & 0.663 & 2.93 & 0.576 \\
dots.tts chosen-only SFT & 8.38 & 72.4 & 0.698 & 2.98 & 7.88 & 67.4 & 0.643 & 2.82 & 0.568 \\
dots.tts ours: AR-16, CP & 4.47 & 76.5 & 0.733 & 3.08 & 6.19 & 69.5 & 0.665 & 2.97 & 0.648 \\
\midrule
Ming base & 30.56 & 44.3 & 0.422 & 2.74 & 12.36 & 43.3 & 0.393 & 2.70 & 0.584 \\
Ming ARDM-DPO (AR-16) & 31.33 & 44.1 & 0.427 & 2.73 & 13.46 & 44.4 & 0.396 & 2.69 & 0.635 \\
Ming chosen-only SFT & 24.33 & 47.1 & 0.441 & 2.72 & 12.56 & 43.2 & 0.386 & 2.61 & 0.548 \\
Ming ours: Diff.-8, NP & 31.61 & 44.0 & 0.425 & 2.74 & 13.66 & 43.7 & 0.394 & 2.73 & 0.622 \\
Ming ours: AR-16, CP & 5.84 & 57.0 & 0.509 & 3.04 & 4.53 & 44.7 & 0.390 & 2.91 & 0.568 \\
Ming ours: AR-16, NP & 5.74 & 58.3 & 0.534 & 2.95 & 4.94 & 47.9 & 0.419 & 2.79 & 0.618 \\
Ming ours: Joint 16{+}8, NP & 5.59 & 58.4 & 0.533 & 2.97 & 5.05 & 48.0 & 0.426 & 2.80 & 0.621 \\
\bottomrule
\end{tabular*}
\end{minipage}
\setcounter{table}{3}

\vspace{7pt}
}]
\subsection{Choosing the flow time}
\label{ssec:time}

Flow time determines the noise level at which the model compares the two candidates. We vary this choice while updating the last eight backbone layers (Table~\ref{tab:train}). Fixed \(t=0.3\) gives the lowest training loss (Fig.~\ref{fig:train}a), but does not give the largest reward gain and reduces speaker similarity relative to the base model. Fixed \(t=0.45\) gives a reward gain of \(+0.137\), with a confidence interval above zero. Full-range \(U(0,1)\) sampling gives a negative reward point estimate (\(-0.069\)), with a confidence interval that includes zero. These estimates do not establish a clear best time without a direct comparison of the reward differences.

These results show why a lower preference loss is insufficient to choose \(t\): different times define different prediction tasks. A model may fit one noise level more easily without improving full-utterance generation. The choice must therefore also be checked on generated speech. The main and module-ablation experiments use discrete schedule times in \([0.3,0.6]\), distinct from the continuous-window ablation in the table.

\subsection{Choosing which module to update}
\label{ssec:module}

We next use the same 687 NP preference pairs and discrete time distribution to compare trainable modules. Updating the last 16 backbone layers gives the largest reward point estimate, \(+0.737\), while reducing WER from \(0.143\) to \(0.060\) and increasing speaker similarity from \(0.466\) to \(0.508\). Updating eight backbone layers gives a smaller improvement; updating only the head remains close to the base model.

Joint 16{+}8 training achieves a lower training loss than backbone-only training (Fig.~\ref{fig:train}b), but its reward gain (\(+0.701\)) is not larger. Thus, updating the conditions supplied to a fixed head is effective in this setting, even though the head produces the acoustic patches. This supports including the backbone in preference post-training. The comparison does not imply that head updates are generally unnecessary, since the update choices also differ in trainable capacity and optimization cost.

\subsection{Candidate source and transfer}
\label{ssec:dots}

Table~\ref{tab:cv3} compares systems on the official English-650 lists~\cite{du2024cosyvoice}. The initialization-only baseline gives limited gains, whereas AR-16 NP/CP variants substantially improve intelligibility. This is consistent with more useful supervision from perturbations beyond initialization and with selective flow-time sampling. For Ming AR-16, NP and CP differ only in candidate perturbation: NP gives higher speaker and emotion similarity, while CP gives higher OVRL and lower emotion-split WER. For the ODE-only dots.tts head, CP improves WER and OVRL on both splits; NP is not applicable. These trade-offs suggest that the perturbation should match the model and target attributes. Because the baseline also differs in time sampling, the comparison does not isolate candidate construction alone.

\section{Summary of findings}
\label{sec:disc}

Sampling changes affect current patches and future conditions, motivating full-utterance preferences. Input-dependent rankings provide supervision that a single fixed inference configuration cannot capture. This supervision benefits backbone adaptation even with a fixed head. Candidate construction should match the supported samplers, while trainable modules should be selected through ablation. Our evidence is based on automatic rewards; listener preferences remain to be evaluated.

\section{Conclusion}
\label{sec:conc}

We studied preference construction and optimization for continuous autoregressive SLMs. Controlled sampling interventions motivate full-utterance comparisons, and a flow-matching preference loss enables effective backbone adaptation with a fixed acoustic head. Time and module ablations show that lower training loss need not yield better speech, highlighting the need to validate post-training choices through generation. Preference updates should therefore account for both acoustic sampling and autoregressive context. Future work will examine longer utterances and validate these gains through listener judgments.

\clearpage
\section{Compliance with Ethical Standards}
This study used publicly available speech datasets, involved no new data collection from human or animal subjects, and did not require ethical approval.

\section{Acknowledgments}
This work is partially supported by the National Natural Science Foundation of China under Grant 62376237,
and the Program for Guangdong Introducing Innovative and Entrepreneurial Teams (Grant No. 2023ZT10X044).

% References follow the ethics statement and acknowledgments on page five.
\begingroup
\small
\bibliographystyle{IEEEbib-six}
\bibliography{strings,ref_v1}
\endgroup
\end{document}